\documentclass[conference]{IEEEtran}
\IEEEoverridecommandlockouts
\usepackage{cite}
\usepackage{amsmath,amssymb,amsfonts}
\usepackage{algorithmic}
\usepackage{graphicx}
\usepackage{textcomp}
\usepackage{xcolor}
\usepackage{comment}
\def\BibTeX{{\rm B\kern-.05em{\sc i\kern-.025em b}\kern-.08em
    T\kern-.1667em\lower.7ex\hbox{E}\kern-.125emX}}

\begin{document}

\title{DeepSaucer: Unified Environment for Verifying Deep Neural Networks
}

%\author{\IEEEauthorblockN{Naoto Sato, Hironobu Kuruma, Yuichiroh Nakagawa, and Hideto Ogawa}
%\IEEEauthorblockA{\textit{Research \& Development Group, } 
%\textit{Hitachi, Ltd.} \\
%Yokohama, Kanagawa, Japan \\
%naoto.sato.je@hitachi.com}
%}

\author{
\IEEEauthorblockN{Naoto Sato\IEEEauthorrefmark{1}, Hironobu Kuruma\IEEEauthorrefmark{1}, Masanori Kaneko\IEEEauthorrefmark{1}, Yuichiroh Nakagawa\IEEEauthorrefmark{1}, Hideto Ogawa\IEEEauthorrefmark{1}, \\
 Thai Son Hoang\IEEEauthorrefmark{2}, and Michael Butler\IEEEauthorrefmark{2}}
\IEEEauthorblockA{\IEEEauthorrefmark{1}Research \& Development Group, Hitachi, Ltd.}
\IEEEauthorblockA{\IEEEauthorrefmark{2}School of Electronics and Computer Science, University of Southampton}
}

\maketitle

\begin{abstract}
In recent years, a number of methods for verifying DNNs have been developed. Because the approaches of the methods differ and have their own limitations, we think that a number of verification methods should be applied to a developed DNN. 
%However, implementation of these methods often depends on the implementation environment, such as machine-learning framework and package libraries. 
%Moreover, the DNN to be verified should be loaded in the same environment as that of the verification method. 
To apply a number of methods to the DNN, it is necessary to translate either the implementation of the DNN or the verification method so that one runs in the same environment as the other. Since those translations are time-consuming, a utility tool, named DeepSaucer, which helps to retain and reuse implementations of DNNs, verification methods, and their environments, is proposed. In DeepSaucer, code snippets of loading DNNs, running verification methods, and creating their environments are retained and reused as software assets in order to reduce cost of verifying DNNs. The feasibility of DeepSaucer is confirmed by implementing it on the basis of Anaconda\textregistered, which provides virtual environment for loading a DNN and running a verification method. In addition, the effectiveness of DeepSaucer is demonstrated by usecase examples. 
% are also shown to help understanding of the effectiveness of DeepSaucer.

\end{abstract}

%\begin{IEEEkeywords}
%machine learning, neural networks, regression analysis 
%\end{IEEEkeywords}

\section{Introduction}

Machine-learning technologies are being gradually introduced in various industrial fields. Among them, deep neural networks (DNNs) are being popularly applied. 

If DNNs are used in safety-critical applications, their behaviors should be carefully verified from several perspectives. In recent years, a number of methods for verifying DNNs have been developed. For example, metamorphic testing \cite{metamor1}\cite{metamor2}\cite{metamor3}\cite{metamor4} is one useful way to evaluate the execution results of DNNs in the case that a test oracle does not exist. In metamorphic testing, metamorphic relations (which are necessary properties of a DNN in relation to multiple input values and their expected output values) are used as pseudo oracles. As another example, neuron-coverage testing \cite{deeptest} \cite{deepxplore} \cite{oxford_mcdc} focuses on activation of neurons in a DNN. Test cases are collected or generated so as to activate neurons that were not activated in the previous testing. One of the advantages of neuron-coverage testing is that it can be applied systematically; that is, it is not necessary to find certain properties (like metamorphic relations) depending on the specification of the DNN . Moreover, a lot of works on formal verification of DNNs by using  SMT (satisfiability modulo theories) or LP (linear programming) solvers have been reported \cite{oxford_smt}\cite{stanford_lp1}\cite{stanford_lp2}\cite{stanford_lp3}\cite{planet}\cite{unified}. The basic notion of formal verification of DNNs is encoding a DNN and its necessary property  as  a logical formula with the theory of  real arithmetic. Solving that formula indicates whether the property is  satisfied.

The authors do not believe that one method is enough for assuring the behavior of DNNs because the approaches of each method are different and have limitations. Therefore, we think that a number of verification methods, such as metamorphic testing and formal verification with a SMT solver, should be applied  to a developed DNN. However, implementation of these methods often depends on the implementation environment, such as versions of Python (e.g., Python2 and Python3), the machine-learning framework (e.g., Tensorflow\texttrademark and Chainer\textregistered), and the package libraries. Moreover, the DNN to be verified should be loaded in the same environment as that of the verification method. Thus, to apply a number of methods to the DNN, it is necessary to translate either the implementation of the DNN or the verification method so that one runs in the same environment as that of the other. As well as DNNs, the datasets used for testing must also be transformed to be consistent with the implementation of the DNN and the verification method. 

Since those translations and transformations are time-consuming, it would be useful if the implementations could be retained and reused for future development. Therefore, we provide a utility tool named {\it DeepSaucer}%\footnote{Available from http://xxxxx.xxxxx.xxxx}
, which helps to retain and reuse implementations of DNNs, verification methods, datasets, and their environments.

\section{Concepts}\label{concepts}
%[Introductionでの説明]手法がたくさんある→実行環境が異なる→実行環境を一致させるためのコード変更が必要→時間がかかるので再利用を促進するツールを提供

When a trained model of a DNN (simply called “model” hereafter) is verified, code including procedures for loading the trained model, loading the dataset to be used for verification, and running a verification function are usually developed together. Among the procedures, those for loading the dataset and running the verification function can be reused for checking a different model. In addition, the procedure for loading the model can be reused for checking the same model with a different dataset or with a different verification function. Therefore, DeepSaucer retains code snippets of these procedures separately as software assets. In this way, it promotes reuse of the procedures and makes it possible to reduce the cost of the verification. Moreover, a lot of deep-learning frameworks and package libraries are publicly available, and they are often updated. As a result, a retained code snippet may run in different environments, such as versions of Python, machine-learning frameworks, and package libraries. When it is necessary to run code snippets in different environments, it is necessary to translate them so that they run in the same environment. Accordingly, similar code snippets based on different environments are retained in DeepSaucer as completely different snippets even if they have the same specification except for their environments.

Moreover, to execute certain code snippets, it is necessary to set up corresponding environments. For example, if version 1.9 of Tensorflow\texttrademark\ is required to run those code snippets, but version 1.8 is installed in the current environment, it is necessary to upgrade to 1.9, and resolve conflicts if they occur. Therefore, as for DeepSaucer, scripts to automatically set up environments for the code snippets are also retained and reused as software assets. It is thus possible to reduce the cost of building the required environment. 

Since environmental requirements are usually defined in "README" in most software packages, they are sometimes ambiguous or insufficient depending on how much detail the developer writes them in "README." However, when DeepSaucer is used, the environmental requirements are defined explicitly and completely as a script. The key concepts of DeepSaucer are listed as follows:
\begin{enumerate}
\item Code snippets of loading trained models, loading datasets, and running verification functions are retained and reused as software assets in order to reduce cost of verifying DNNs
\item Scripts to create environments automatically are also retained and reused to prevent the environmental requirements from being ambiguous or insufficient and to reduce cost of building required environments.
\end{enumerate}

\section{Implementation}

\begin{figure}[htbp]
\scalebox{0.39}{\includegraphics[bb=0 0 618 370]{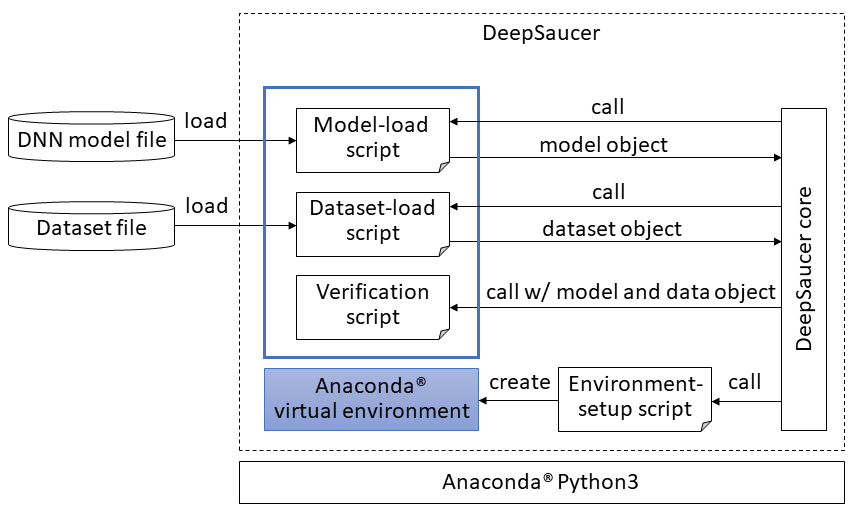}}
\caption{Software architecture of DeepSaucer}
\label{fig01}
\end{figure}

DeepSaucer is implemented on the basis of Python 3.6 provided by Anaconda\textregistered. The architecture of DeepSaucer is shown in Fig. \ref{fig01}.

As code snippets mentioned in Section \ref{concepts}, DeepSaucer retains Python scripts for loading models, loading datasets for testing, and executing verification functions. They are called {\it model-load script}, {\it dataset-load script}, and {\it verification script} respectively. A user selects one of them to be executed. A model-load script is called by the DeepSaucer core. Typically, a model-load script refers to a particular file containing information about a trained model. Then, the model-load script returns a trained DNN model to DeepSauce core. Similarly, a dataset-load script is called by the DeepSaucer core. The dataset-load script  returns the corresponding dataset that  was transformed appropriately. After running the model-load script and dataset-load script, a verification script  is called by the DeepSaucer core. The verification script executes a particular verification function with the loaded model and dataset.

\begin{figure}[htbp]
\scalebox{0.45}{\includegraphics[bb=0 0 528 441]{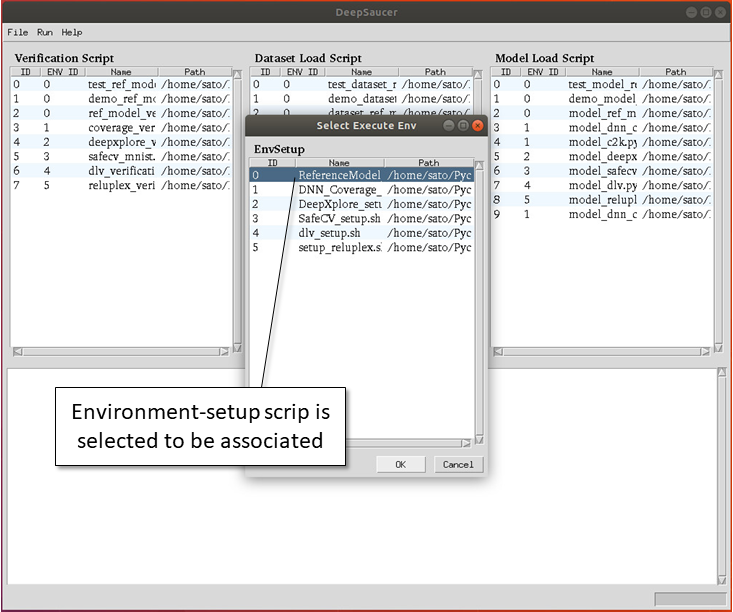}}
\caption{Screen shot of DeepSaucer in  while it is  associating a model-load script with an environment-setup script}
\label{fig02}
\end{figure}

\begin{figure}[htbp]
\scalebox{0.45}{\includegraphics[bb=0 0 528 441]{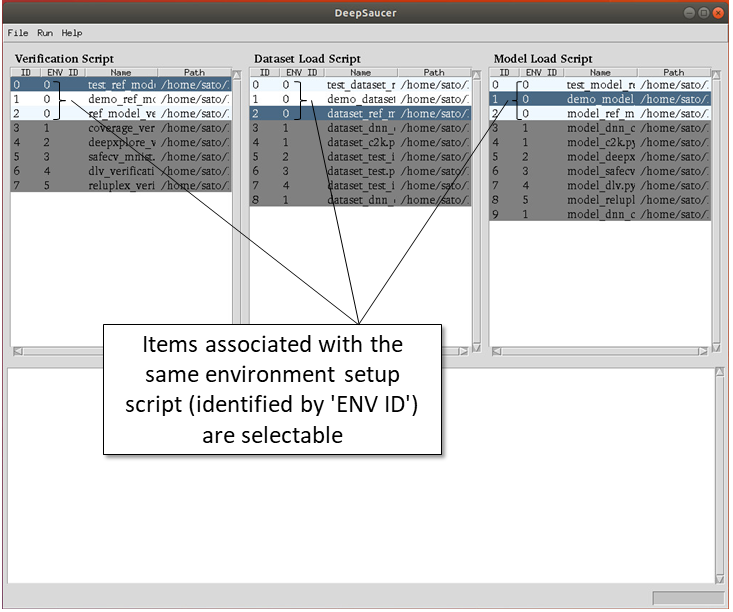}}
\caption{Screen shot of DeepSaucer in  while it is selecting a model-load script, a dataset- load script, and a verification script}
\label{fig03}
\end{figure}

\begin{figure}[htbp]
\scalebox{0.45}{\includegraphics[bb=0 0 528 441]{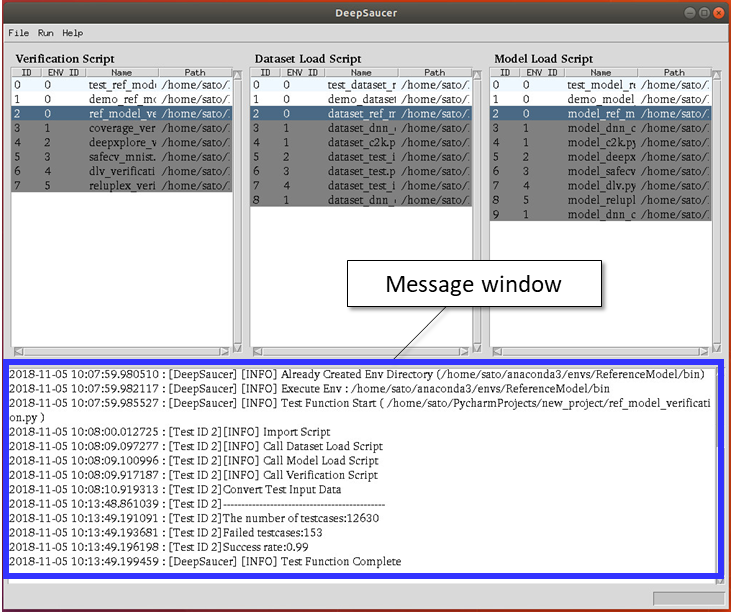}}
\caption{Screen shot of DeepSaucer showing the result of verification in the bottom window}
\label{fig04}
\end{figure}

The model-load script, dataset-load script, and verification script (hereafter, all called “functional scripts”) are basically expected to written in Python\footnote{When it is necessary to use an executable program developed in other programming languages, it is possible to write a Python script in which the executable program is called}. However, it is allowed that a functional script requires a different environment. When a functional script is loaded in DeepSaucer, it is associated with an {\it environment-setup script} by a user (Fig. \ref{fig02}); consequently,  an Anaconda\textregistered\ virtual environment is created, and necessary package libraries appropriate for the corresponding functional script are installed in that environment. That is, environment-setup scripts include the commands "conda", "pip", and so on. The model-load script, dataset-load script, and verification script can be selected for running the verification only if they are associated with the same environment-setup script (Fig. \ref{fig03}). Before the DeepSaucer core runs the selected functional scripts, it calls the associated environment-setup script. In the Anaconda\textregistered\ virtual environment built by the environment-setup script, the selected functional scripts are executed. The environment-setup script  are written in Shell Script on Linux or Batch Script on Windows, and they are expected to be loaded in DeepSaucer before the functional scripts are loaded. After running the verification, the result is shown in the bottom window of the DeepSaucer screen  (Fig. \ref{fig04}). At the moment, it is assured that DeepSaucer runs on Ubuntu 17.10 with Python 3.6 provided by Anaconda\textregistered\ 5.2.

%In most of the machine learning frameworks, a DNN can be saved as a model data file. However, the saved model data file often contains only the parameters of the DNN such as weight and bias. So, when we want to load the DNN from the model data file, we should compose the structure of the DNN before loading it. The structure of the DNN is defined as the number of layers, type of each layer, the number of neurons in each layer, and so on. We think necessary assets of DNNs are not only their parameters, but also their structures. Therefore, DeepSaucer does not keep model files but executable python scripts in which both the structure and the parameter is set up as assets. Similarly, dataset for testing is kept as a script in which the dataset is loaded and trasformed into a particular format.
%モデルはパラメータのデータだけでなく，レイヤ構造やノード数，活性化関数などの情報が必要．それらの情報を構成して，モデルをビルドするスクリプトを資産として保存する．
%⇒Kerasのモデルは，パラメータだけでなくレイヤ構造なども保存できるらしい
%テストに使用するデータも同様で，データをロードして適切に加工するスクリプトを資産として保存する

\section{Use-case examples}
\subsection{Reuse of verification script}\label{usecase1}

In this use-case example, it is assumed that a DNN model was developed and it is necessary to check it by several verification methods. DeepSaucer retains a number of verification scripts and the corresponding datasets used in past projects. It is also assumed that Chainer\textregistered\ on Python 2.7 was chosen in the current project (although Tensorflow\texttrademark\ on Python 3.6 was adopted in the past projects). In that case, verification scripts retained in DeepSaucer is not applicable to the current model directly. Therefore, parameters (such as weight and bias) of the model developed in Chainer\textregistered\ based on Python 2.7 are first saved as a file. Second, a model-load script that reads the saved file and returns the model of Tensorflow\texttrademark\ is developed on Python 3.6. Finally, by running the existing verification scripts with the model , the model is successfully verified. 

In such a situation, DeepSaucer promotes the reuse of existing verification scripts. Moreover, since it shows a list of verification scripts used in the past projects, which verification scripts are reusable is easy to understand. Similarly, since DeepSaucer presents the environments of the verification scripts, how the current model should be translated is also easy to understand.

\subsection{Reuse of environment-setup script}\label{usecase2}

In regard to almost the same situation as that described in Section \ref{usecase1},  it is additionally supposed that the environmental requirements in the past projects are all different. In that case, to reuse the existing verification scripts, the same number of environments as that in the past projects needs to be built. However, since DeepSaucer retains environment-setup scripts, it is possible to automatically create the environments required to run the existing verification scripts.

\section{Conclusion}
We  have developed a utility tool, called DeepSaucer, for verifying DNNs. DeepSaucer helps to retain and reuse implementations of DNNs, verification methods, datasets, and their environments. The key concepts of DeepSaucer are summarized as follows. Code snippets of loading trained models, loading datasets, and running verification functions are retained and reused as software assets. As a result, it is possible to reduce the cost of verifying DNNs. To prevent the environmental requirements from being ambiguous or insufficient and to reduce the cost of building required environments, scripts to create environments automatically are also retained and reused. The feasibility of DeepSaucer is confirmed by implementing it on the basis of Python 3.6 provided by Anaconda\textregistered. In addition, the effectiveness of DeepSaucer is demonstrated by the usecase examples. 
As for future work, the effectiveness of DeepSaucer will be evaluated in actual developments of DNNs. Moreover, our implementation using Anaconda\textregistered will be compared with other implementations based on other virtualization tools like Docker\textregistered.

%%%%%%%%%%%%%%%%%%%%%%%%%%%%%%%%%%%%%%%%%%%%%%%%%%%%%%%%%%%%%%%%%%%%%%%%%%%%%%%%%%%%%%%%%%%%%%%%

%\section*{References}

\end{document}